\documentclass[useAMS,usenatbib]{mn2e}
\usepackage{graphicx,amsmath}
\usepackage[papersize={8.5in,11in}]{geometry}
\renewcommand{\vec}[1]{{\bf{#1}}}

\begin{document}

\title{Modified gravity and the origin of inertia}

\author[J. W. Moffat and V. T. Toth]{J. W. Moffat$^{1,2}$ and V. T. Toth$^1$\\
$^1$Perimeter Institute for Theoretical Physics, Waterloo, Ontario N2L 2Y5, Canada\\
$^2$Department of Physics, University of Waterloo, Waterloo, Ontario N2L 3G1, Canada}

\maketitle

\volume{395}
\pagerange{L25--L28}
\pubyear{2009}

\begin{abstract}
Modified gravity theory is known to violate Birkhoff's theorem. We explore a key consequence of this violation, the effect of distant matter in the Universe on the motion of test particles. We find that when a particle is accelerated, a force is experienced that is proportional to the particle's mass and acceleration and acts in the direction opposite to that of the acceleration. We identify this force with inertia. At very low accelerations, our inertial law deviates slightly from that of Newton, yielding a testable prediction that may be verified with relatively simple experiments. Our conclusions apply to all gravity theories that reduce to a Yukawa-like force in the weak field approximation.
\end{abstract}

\begin{keywords}
gravitation - relativity.
\end{keywords}

\section{Introduction}

Modified Gravity (MOG; \cite{Moffat2006a}) is a fully relativistic theory of gravitation that is based on postulating the existence of a massive vector field, $\phi_\mu$. The choice of a massive vector field is motivated by our desire to introduce a {\em repulsive} modification of the law of gravitation at short range. The vector field is coupled universally to matter. The theory, therefore, has three constants: in addition to the gravitational constant $G$, we must also consider the coupling constant $\omega$ that determines the coupling strength between the $\phi_\mu$ field and matter, and a further constant $\mu$ that arises as a result of considering a vector field of non-zero mass that controls the coupling range. The theory promotes $G$, $\omega$, and $\mu$ to scalar fields, hence they are allowed to run with distance or energy.

MOG is consistent with observation ranging from the scale of the solar system \citep{Moffat2007e} to globular clusters \citep{Moffat2007a}, galaxy rotation curves \citep{Moffat2004,Brownstein2006a,Moffat2007e}, galaxy cluster masses \citep{Brownstein2006b}, gravitational lensing \citep{Moffat2008a} of the Bullet Cluster 1E0657-558 \citep{Brownstein2007}, and cosmological observations \citep{Moffat2007c}, without free parameters.

General relativity satisfies Birkhoff's theorem \citep{BIRKHOFF1923}. The metric inside an empty spherical cavity in the center of a spherically symmetric system is the Minkowski metric \citep{WEINBERG1972}. Distant matter does not influence the equations of motion of particles within the spherical cavity. This is the basic difficulty of attempts, such as that of \cite{SCIAMA1953}, to deduce Mach's principle from the postulates of general relativity. That Mach's principle and general relativity may not be compatible with each other is also demonstrated by \cite{BRANS1962a}.

In contrast, MOG is known to violate Birkhoff's theorem. Inside a spherically symmetric shell of matter, the MOG force is non-zero.
In the present paper, we investigate the consequences of this fact.

\section{The law of inertia}

\begin{figure}
\centering
\includegraphics[width=0.75\linewidth]{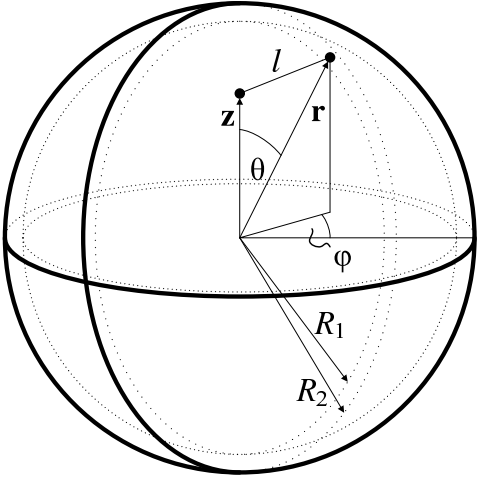}
\caption{Parameterizing a thin spherical shell and a test particle within.}
\label{fig:shell}
\end{figure}

We investigate MOG in the weak-field approximation, in the case of a spherically symmetric, homogeneous shell of density $\rho$, inner radius $R_1$ and outer radius $R_2$, with a test particle of mass $m$ located at $\vec{z}$, at distance $z$ from the shell's center (Figure~\ref{fig:shell}). We parameterize a point $\vec{r}$ on the shell using spherical coordinates $r$, $\theta$, and $\phi$, where $\theta$ is the angle between the line connecting a point in the shell with the center of the shell, and the line connecting the test particle to the center of the shell, and $\phi$ is the angle of rotation in a plane perpendicular to the line connecting the test particle to the center of the shell, relative to a preferred direction.

The distance $l$ between a point in the shell and the test particle can be written as
\begin{equation}
|\vec{r}-\vec{z}|^2=l^2=z^2+r^2-2rz\cos{\theta}.
\end{equation}
In the Newtonian case, the gravitational force obeys the inverse square law. The gravitational force on the test particle can therefore be written as
\begin{equation}
\vec{F}=\int \frac{G\rho m(\vec{r}-\vec{z})}{l^3}~dV,
\label{eq:F}
\end{equation}
where $G=G_N$ is Newton's constant, $dV$ denotes a volume element inside the shell, and the integration is carried out over the entire volume of the shell. The volume element can be expressed using coordinates as
\begin{equation}
dV=r^2\sin{\theta}~dr~d\theta~d\phi.
\end{equation}

Because of spherical symmetry, components of $\vec{F}$ perpendicular to the line connecting the test particle and the center of the shell vanish. The component parallel to this line, which we denote with $F_0$, can be calculated using the projection coefficient $(r\cos{\theta}-z)/l$:
\begin{equation}
F_0=\int\limits_0^{2\pi}\int\limits_0^\pi\int\limits_{R_1}^{R_2}\frac{G_N\rho mr^2(r\cos{\theta}-z)\sin{\theta}}{l^3}~dr~d\theta~d\phi.
\end{equation}
This integral evaluates to
\begin{equation}
F_0=\begin{cases}
0&z<R_1\\
4\pi G_N\rho m(R_1^3-z^3)/3z^2&R_1\leq z<R_2\\
4\pi G_N\rho m(R_1^3-R_2^3)/3z^2&R_2\leq z.\\
\end{cases}
\end{equation}

In the weak-field approximation, the MOG acceleration law can be written as Newton's law of gravity with an effective gravitational constant that incorporates a repulsive Yukawa term\footnote{Setting $G_\infty=(1+\alpha)G_N$ we recover Newton's law of gravity in the limit $l\rightarrow 0$, as shown by~\cite{Moffat2006a}.}:
\begin{equation}
G=G_\infty\left[1-\frac{\alpha}{1+\alpha}\left(1+\mu l\right)e^{-\mu l}\right],
\end{equation}
where $\alpha$ controls the strength of the Yukawa contribution, and $\mu$ is a range parameter. In the following we use only this weak field approximation. Therefore, our conclusions may be applicable to theories other than MOG, so long as in the weak field approximation, they also yield a Yukawa-like modification of gravity.

In the cosmological context, the coefficient $\mu$ is set to the reciprocal of the horizon scale $c/H_0$, where $c$ is the speed of light and $H_0$ is Hubble's constant:
\begin{equation}
\mu=\frac{H_0}{c},
\label{eq:mu}
\end{equation}
which yields good agreement with key cosmological observations \citep{Moffat2007c}.

In the case of a spherical volume of uniform mass density and radius $R$, after evaluating ($\ref{eq:F}$), the gravitational force in the interior and the exterior, denoted by $F_1$ and $F_2$, respectively, are written as
\begin{align}
&F_1(R,z)=-\pi G_\infty\rho m\left(\frac{4z}{3}-\frac{2\alpha}{1+\alpha}\right.\\
&\left.\times\frac{(\mu R+1)\left[(\mu z+1)e^{-\mu(R+z)}+(\mu z-1)e^{-\mu(R-z)}\right]}{\mu^3z^2}\right),\nonumber\\
&F_2(R,z)=-\pi G_\infty\rho m\left(\frac{4R^3}{3z^2}-\frac{2\alpha}{1+\alpha}\right.\\
&\left.\times\frac{(\mu z+1)\left[(\mu R+1)e^{-\mu(z+R)}+(\mu R-1)e^{-\mu(z-R)}\right]}{\mu^3z^2}\right).\nonumber
\end{align}
The gravitational force inside, within, and outside a spherical shell of inner radius $R_1$ and outer radius $R_2$ can be written as
\begin{equation}
F_0=\begin{cases}
F_1(R_2,z)-F_1(R_1,z)&z<R_1\cr
F_1(R_2,z)-F_2(R_1,z)&R_1\leq z<R_2\cr
F_2(R_2,z)-F_2(R_1,z)&R_2\leq z.\cr
\end{cases}
\end{equation}

In particular, $F_1(R_2,z)-F_1(R_1,z)$ is nonvanishing; the net force in the interior of a spherically symmetric shell is not zero.

For an infinitesimally thin shell, we can write:
\begin{equation}
dF=\lim_{R_2\rightarrow R_1}\left[F_1(R_2,z)-F_1(R_1,z)\right],
\end{equation}
or
\begin{equation}
{\cal F}(R,z)=\frac{dF}{dR}=\left.\lim_{R_2\rightarrow R_1}\frac{F_1(R_2,z)-F_1(R_1,z)}{R_2-R_1}\right|_{R_1=R}.
\end{equation}

We imagine a test particle that is surrounded by an infinite series of infinitesimally thin concentric shells of matter. Located at the center of the shells, the test particle experiences no net force. We now bring the particle into motion with velocity $v$, and the shells are dragged along. However, because of the finite propagation velocity of the gravitational interaction, the shells will no longer appear concentric as viewed from the vantage point of the moving particle. Therefore, a force will act on the particle. This force can be calculated by integrating the displacement of the particle relative to each of the concentric shells while taking into account the particle's motion:
\begin{equation}
F(z,v)=\int\limits_0^\infty{\cal F}\left(R,\frac{cz+vR}{c+v}\right)~dR.
\label{eq:FF}
\end{equation}
This expression allows us to calculate the displacement $z$ associated with velocity $v$ at which forces are in equilibrium and the net force acting on the particle is zero, by solving the equation
\begin{equation}
F(z,v)=0
\end{equation}
for $z$. This equation is difficult to solve exactly, but for nonrelativistic velocities, it yields
\begin{equation}
z\simeq-\frac{2}{\mu c}v.
\end{equation}
Differentiating this equation with respect to $t$ gives
\begin{equation}
\frac{dz}{dt}\simeq-\frac{2}{\mu c}\frac{dv}{dt}=-\frac{2}{\mu c}a.
\label{eq:dzdt}
\end{equation}

The value $dz/dt$ is a velocity, specifically the velocity with which the equilibrium location moves relative to the accelerating particle.

Assuming that the particle remains in equilibrium, we substitute (\ref{eq:dzdt}) back into equation~(\ref{eq:FF}), to get
\begin{align}
F(a)=&\lim_{z\rightarrow 0}F\left(z,-\frac{dz}{dt}\right)=\frac{\pi G_\infty\rho m}{2\mu^2c^2}\frac{\alpha}{1+\alpha}\label{eq:F1}\\
\times&\frac{(\mu c^2+2a)^2\left[\mu c^2\log{\left(1+\frac{4a}{\mu c^2}\right)}-4a\frac{\mu c^2+2a}{\mu c^2+4a}\right]}{a^2}.\nonumber
\end{align}
When $a\gg\mu c^2$, this expression can be simplified. In particular, the linear term inside the square brackets will dominate over the logarithmic term, which can therefore be omitted. We are left with
\begin{align}
F(a)\simeq&-\frac{\pi G_\infty\rho m}{2\mu^2c^2}\frac{\alpha}{1+\alpha}\frac{4(\mu c^2+2a)^3}{a(\mu c^2+4a)}\simeq\nonumber\\
&-\frac{4\pi G_\infty\rho}{\mu^2c^2}\frac{\alpha}{1+\alpha}ma.
\label{eq:Fa}
\end{align}

Given that $G_\infty=(\alpha+1)G_N$ and having assumed $\mu c=H_0$ (\ref{eq:mu}), we can rewrite equation~(\ref{eq:Fa}) as
\begin{equation}
F(a)\simeq -\frac{3}{2}\Omega_M\alpha ma,
\label{eq:F2}
\end{equation}
where $\Omega_M=\rho/\rho_\mathrm{crit}$ represents the matter content (associated with $\rho$) of the universe, while $\rho_\mathrm{crit}=3H_0^2/8\pi G_N$ is the critical density.

Given $\Omega_M\simeq 0.05$ (assuming no exotic dark matter, an assumption that is consistent with cosmological observations when using the MOG cosmology \citep{Moffat2007c}), if we use $\alpha\simeq 13.3$, we get
\begin{equation}
F(a)\simeq -ma.
\label{eq:F3}
\end{equation}
This is formally identical to Newton's law of inertia. In Newton's theory, the force
\begin{equation}
F=ma
\end{equation}
accelerating the particle with acceleration $a$ is counteracted by an inertial force $F_I$ of equal magnitude:
\begin{equation}
F_I=-ma,
\end{equation}
such that the sum of the two is zero,
\begin{equation}
F_I+F=0,
\end{equation}
in accordance with the principle set forth by \cite{DALEMBERT1743}.

In our case, we made no {\it a priori} assumption about the existence of inertia, but we recovered a force $F(a)$ acting on a particle due to the presence of distant matter that plays the same role as d'Alembert's inertial force $F_I$. The force $F(a)$ arose as a result of the influence of distant matter in the universe on the test particle, offering an effective realization of Mach's principle \citep{MACH1883}.

We must emphasize that this law was recovered, in the weak field approximation of our modified gravity, from the gravitational acceleration alone; inertia was not postulated either explicitly or implicitly.

The mass $m$ on the right-hand side of equation~(\ref{eq:F3}) is the {\em passive gravitational mass}, characterizing how an object responds to an external gravitational field. We can also define the {\em inertial mass} of an object as
\begin{equation}
m_I=-\frac{F(a)}{a}.
\end{equation}
The equivalence of the passive gravitational and inertial mass is often referred to as the weak equivalence principle.

\begin{figure}
\includegraphics[width=\linewidth]{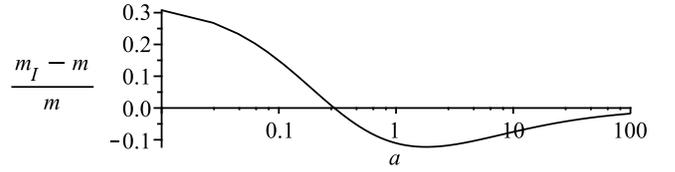}
\caption{Does MOG violate the weak equivalence principle for very small accelerations? The horizontal axis in this plot is acceleration, measured in units of the ``cosmic acceleration'' $cH_0\simeq 7\times 10^{-10}$~m/s$^2$. The vertical axis shows the difference between inertial mass $m_I=-F(a)/a$ and passive gravitational mass $m$, as predicted by equation~(\ref{eq:F1}).}
\label{fig:equiv}
\end{figure}

\begin{figure}
\includegraphics[width=\linewidth]{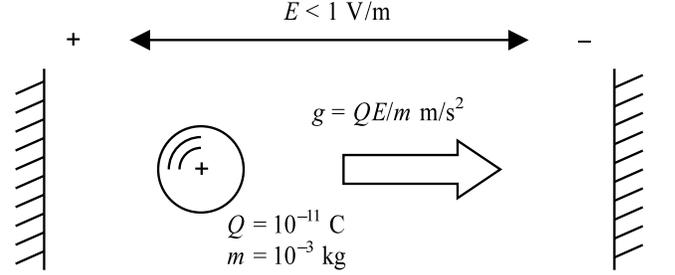}
\caption{Schematic of a simple experiment that can be used to verify the validity of the force law $F=ma$ for very small accelerations. With the values presented here, a measurement of a deflection of $\sim 1.8$~mm over the course of ten minutes with an accuracy better than $\sim$10\% is required, in order to measure the deficit in inertial mass. A smaller acceleration (corresponding to $E\simeq 0.01$~V/m) could be used to measure an excess in inertial mass of up to $\sim$30\%.}
\label{fig:experiment}
\end{figure}

For {\em very} small accelerations ($a\ll cH_0$), the approximation that led to equation~(\ref{eq:F3}) no longer applies. Instead, we can use $\log{(1+x)}=x-x^2/2+x^3/3+{\cal O}(x^4)$ (we need to expand to $x^3$, as the first two terms are canceled out in equation~\ref{eq:F1}), to get
\begin{equation}
F(a)\simeq-\frac{16\pi G_\infty\rho}{3\mu^2c^2}\frac{\alpha}{1+\alpha}ma=-\frac{4}{3}ma.
\label{eq:Fma}
\end{equation}

This is a profound result: we are, in fact, predicting a small violation of the weak equivalence principle for accelerations $a\sim cH_0$ (see Fig.~\ref{fig:equiv}.) This is a testable prediction, especially in view of the fact that the acceleration $a$ in equation~(\ref{eq:F1}) does not need to be gravitational in origin.

\section{Discussion}

Fig.~\ref{fig:experiment} depicts schematically a simple experiment that could be used to verify the validity of Newton's law of inertia at very small accelerations. The values of masses, lengths, time scales, and charges are not extreme in nature. Indeed, an experiment of this nature was performed\footnote{We thank Dr. Ignacio Trujillo from the {\em Instituto de Astrofísica de Canarias} for bringing this paper to our attention.} in 1986 \citep{AV1986}, and confirmed Newton's second law at accelerations as low as $\sim 3\times 10^{-11}$~m/s$^2$. However, this experiment was performed on the surface of the Earth.

Our predicted deviation from Newton's second law takes place when the acceleration of the test mass is very small {\em relative to the distant stars}, to borrow Mach's famous notion. This does not necessarily imply an inertial system; for instance, a particle in orbit around the Earth or the Sun is continuously accelerating with respect to the cosmic microwave background (CMB) even though it is in geodesic motion. Therefore, we believe that the experiment depicted in Fig.~\ref{fig:experiment} must be performed in a space-based laboratory that maintains uniform velocity with respect to the CMB through the use of thrusters. On the other hand, the experiment does not need to rely on exotic technologies, long flight duration, or a flight into deep space, far from the Sun, utilizing costly deep space communication and a nuclear power source.

To compensate for accelerations relative to the CMB, one must take into account the acceleration relative to the Sun, and the solar system's acceleration relative to the CMB. The dominant term in the latter is acceleration toward the galactic center, which can be computed from the parameters of the Sun's galactic orbit. If this acceleration is accounted for, any residual acceleration is probably small enough so that we are already in the regime where (\ref{eq:Fma}) applies. Nonetheless, additional corrections are obtained when one considers the acceleration of the galactic center relative to the CMB. This estimate could proceed in steps, first by estimating the acceleration of the galactic center toward M31, and then estimating the acceleration of the Local Group relative to the CMB\footnote{We thank the anonymous referee for elaborating on this point.}.

An intriguing alternative to a space-based experiment was proposed by \cite{IGN2007}, who noted that specific terrestrial locations at high latitude twice yearly experience brief periods of near zero acceleration relative to the galactic center. This may permit Earth-based tests of effects that are predicted to occur in frames that are not accelerating relative to the galactic center or the CMB.

\newpage

\section*{Acknowledgments}

The research was partially supported by National Research Council of Canada. Research at the Perimeter Institute for Theoretical Physics is supported by the Government of Canada through NSERC and by the Province of Ontario through the Ministry of Research and Innovation (MRI).

\vspace{1em}

\bibliography{refs}

\begin{thebibliography}{}

\bibitem[\protect\citeauthoryear{{Abramovici} \& {Vager}}{{Abramovici} \&
  {Vager}}{1986}]{AV1986}
{Abramovici} A.,  {Vager} Z.,  1986, \prd, 34, 3240

\bibitem[\protect\citeauthoryear{Birkhoff}{Birkhoff}{1923}]{BIRKHOFF1923}
Birkhoff G.,  1923, Relativity and Modern Physics.
Harvard Univ. Press., Cambridge, MA, USA

\bibitem[\protect\citeauthoryear{Brans}{Brans}{1962}]{BRANS1962a}
Brans C.~H.,  1962, Phys. Rev., 125, 388

\bibitem[\protect\citeauthoryear{{Brownstein} \& {Moffat}}{{Brownstein} \&
  {Moffat}}{2006a}]{Brownstein2006b}
{Brownstein} J.~R.,  {Moffat} J.~W.,  2006a, \mnras, 367, 527

\bibitem[\protect\citeauthoryear{{Brownstein} \& {Moffat}}{{Brownstein} \&
  {Moffat}}{2006b}]{Brownstein2006a}
{Brownstein} J.~R.,  {Moffat} J.~W.,  2006b, \apj, 636, 721

\bibitem[\protect\citeauthoryear{{Brownstein} \& {Moffat}}{{Brownstein} \&
  {Moffat}}{2007}]{Brownstein2007}
{Brownstein} J.~R.,  {Moffat} J.~W.,  2007, \mnras, {382 (1)}, 29

\bibitem[\protect\citeauthoryear{d'Alembert}{d'Alembert}{1743}]{DALEMBERT1743}
d'Alembert J.,  1743, Trait\'e de dynamique.
David, Paris

\bibitem[\protect\citeauthoryear{{Ignatiev}}{{Ignatiev}}{2007}]{IGN2007}
{Ignatiev} A.~Y.,  2007, Physical Review Letters, 98, 101101

\bibitem[\protect\citeauthoryear{Mach}{Mach}{1883}]{MACH1883}
Mach E.,  1883, Die Mechanik in ihrer Entwickelung: Historisch-kritisch
  dargestellt.
Leipzig: F. A. Brockhaus

\bibitem[\protect\citeauthoryear{{Moffat}}{{Moffat}}{2004}]{Moffat2004}
{Moffat} J.~W.,  2004, ArXiv, {gr-qc/0404076}

\bibitem[\protect\citeauthoryear{{Moffat}}{{Moffat}}{2006}]{Moffat2006a}
{Moffat} J.~W.,  2006, Journal of Cosmology and Astroparticle Physics, 2006,
  004

\bibitem[\protect\citeauthoryear{{Moffat} \& {Toth}}{{Moffat} \&
  {Toth}}{2007}]{Moffat2007c}
{Moffat} J.~W.,  {Toth} V.~T.,  2007, ArXiv, 0710.0364 [astro-ph]

\bibitem[\protect\citeauthoryear{{Moffat} \& {Toth}}{{Moffat} \&
  {Toth}}{2008}]{Moffat2007a}
{Moffat} J.~W.,  {Toth} V.~T.,  2008, \apj, 680, 1158

\bibitem[\protect\citeauthoryear{{Moffat} \& {Toth}}{{Moffat} \&
  {Toth}}{2009a}]{Moffat2007e}
{Moffat} J.~W.,  {Toth} V.~T.,  2009a, Class. Quant. Grav., 26, 085002

\bibitem[\protect\citeauthoryear{{Moffat} \& {Toth}}{{Moffat} \&
  {Toth}}{2009b}]{Moffat2008a}
{Moffat} J.~W.,  {Toth} V.~T.,  2009b, \mnras, (accepted for publication)

\bibitem[\protect\citeauthoryear{Sciama}{Sciama}{1953}]{SCIAMA1953}
Sciama D.,  1953, \mnras, 113, 34

\bibitem[\protect\citeauthoryear{Weinberg}{Weinberg}{1972}]{WEINBERG1972}
Weinberg S.,  1972, Gravitation and Cosmology.
John Wiley \& Sons

\end{thebibliography}
\bibliographystyle{mn2e}

\end{document}